\documentclass[twocolumn,secnumarabic,amssymb, amsmath,
nofootinbib,tightenlines,  nobibnotes, aps, prl,epsfig]{revtex4}
\usepackage{graphicx}% Include figure files
\usepackage{dcolumn}% Align table columns on decimal point
\usepackage{bm}% bold math
\begin{document}
%\preprint{APS/123-QED}
\title{ The longitudinal structure function  $F_{L}$ from the charm structure function $F_{2}^{c}$ }% Force line breaks with \\
\author{B.Rezaei }
\email{brezaei@razi.ac.ir}%Lines break automatically or can be forced with \\
\author{G.R.Boroun}%
\altaffiliation{grboroun@gmail.com; boroun@razi.ac.ir }
\affiliation{ Physics Department, Razi University, Kermanshah
67149, Iran}% \textbackslash\textbackslash
\date{\today}% It is always \today, today,
\begin{abstract}
We predict the effect of the charm structure function on the
longitudinal structure function at small $x$. In NLO analysis we
find that the hard Pomeron behavior gives a good description of
$F_{L}$ and $F_{k}^{c}(k=2,L)$ at small $x$ values. We conclude
that a direct relation between
$F_{L}{\propto}\hspace{0.1cm}F_{2}^{c}$ would provide useful
information on how to measurement longitudinal structure function
at high $Q^{2}$ values. Having checked that this model gives a
good description of the data, when compared with other
models.\end{abstract}
 %PACS, the Physics and Astronomy
                              %Classification Scheme.
 %Use showkeys class option if keyword
\maketitle
%%%%%%%%%%%%%%%%%%%%%%%%%%%%%%%%%%%%%%%%%%%%%%%%%%%%%%%%%%%%%%%%%
In the LO (leading order), authors in Ref.[1] have suggested an
approximation relation between the gluon density and the
longitudinal structure function $F_{L}$, which demonstrates the
close relation between the longitudinal structure function and the
gluon density. Therefore the longitudinal structure function is  a
very clean probe of the small $x$ gluon distribution. We
specifically consider the next- to- leading- order (NLO)
corrections to the longitudinal structure function $F_{L}$,
projected from the hadronic tensor by combination of the metric
and the spacelike momentum transferred by the virtual photon
$(g_{\mu\nu}-q_{\mu}q_{\nu}/q^{2})$. In the next- to -leading
order the longitudinal structure function is proportional to
hadronic tensor as follows:
\begin{equation}
F_{L}(x,Q^{2})/x=\frac{8x^{2}}{Q^{2}}p_{\mu}p_{\nu}W_{\mu\nu}(x,Q^{2}),
\end{equation}
where $p^{\mu}(p^{\nu})$ is the hadron momentum and $W^{\mu\nu}$
is the hadronic tensor. In this relation we neglecting the hadron
mass.\\

The basic hypothesis is that the total cross section of a hadronic
process can be written as the sum of the contributions of each
parton type (quarks, antiquarks, and gluons) carrying a fraction
of the hadronic total momentum. In the case of deep- inelastic-
scattering it reads:
\begin{equation}
d\sigma_{H}(p)=\sum_{i}{\int}dyd\hat{\sigma}_{i}(yp)\Pi_{i}^{0}(y),
\end{equation}
where $d\hat{\sigma}_{i}$ is the cross section corresponding to
the parton $i$ and $\Pi_{i}^{0}(y)$ is the probability of finding
this parton in the hadron target with the momentum fraction $y$.
Now, taking into account the kinematical constrains one gets the
relation between the hadronic and the partonic structure
functions:
\begin{eqnarray}
f_{j}(x,Q^{2})&=&\sum_{i}{\int}_{x}^{1}\frac{dy}{y}\textsf{f}_{j}(\frac{x}{y},Q^{2})\Pi_{i}^{0}(y)\\\nonumber
&&=\sum_{i}\textsf{f}_{j}{\otimes}\Pi_{i}^{0}(y)\hspace{0.5cm},j=2,L,
\end{eqnarray}
where $\textsf{f}_{j}(x,Q^{2})=F_{j}(x,Q^{2})/x$ and the symbol
${\otimes}$ denotes convolution according to the usual
prescription. Equation (3) expresses the hadronic structure
functions as the convolution of the partonic structure function,
which are calculable in perturbation theory, and the probability
of finding a parton in the hadron which is a nonperturbative
function. So, in correspondence with Eq.(3) one can write Eq.(1)
for the gluon density dominated at low $x$ values by follows:
\begin{eqnarray}
F^{g}_{L}/x{=}\frac{\alpha_{s}}{4\pi}[\textsf{f}_{L,G}^{(1)}{\otimes}g^{0}]+(\frac{\alpha_{s}}{4\pi})^{2}[\textsf{f}_{L,G}^{(2)}{\otimes}g^{0}],
\end{eqnarray}
where $\textsf{f}_{L,G}^{(1)}$ and $\textsf{f}_{L,G}^{(2)}$ are
the LO and NLO partonic longitudinal structure function
corresponding to gluons, respectively [2-3]. We present the
expressions, after full agreement has been achieved, in the form
of kernels $K^{G}$ which give NLO- $F_{L}$ upon convolution with
 the gluon distribution:
\begin{eqnarray}
F^{g}_{L}(x,Q^{2})&=&K^{G}(\frac{x}{y},Q^{2}){\otimes}G(y,Q^{2})\\\nonumber
&&=\int_{x}^{1}\frac{dy}{y}K^{G}(\frac{x}{y},Q^{2})G(y,Q^{2}),
\end{eqnarray}
where $K^{G}(x,Q^{2})$ is the DIS coefficient function and the
explicit form of it$^{,}$s appears in the
 Appendix. \\
Exploiting the small $x$ asymptotic behavior of the gluon
distribution [4] into the symbolic form,
\begin{equation}
G(x,Q^{2})|_{x{\rightarrow}0}{\rightarrow}x^{-\delta},
\end{equation}
where the exponent of the gluon distribution is found to be either
$\delta{\approx}0$ or $\delta{\approx}0.5$. The first value
corresponds to the soft pomeron intercept and the second value to
the hard
(Lipatov) pomeron intercept.\\
Using Eq.6 in 5, the integration of the gluon kernel over
$z=\frac{x}{y}$ and finally summing over the gluon distribution
function $G(x,Q^{2})$ yields:
\begin{eqnarray}
F^{g}_{L}(x,Q^{2})&=&G(x,Q^{2})[K^{G}(1-z,Q^{2}){\otimes}(1-z)^{\delta}]\\\nonumber
&&=G(x,Q^{2})\int_{0}^{1-x} K^{G}(1-z,Q^{2})(1-z)^{\delta} dz.
\end{eqnarray}

The objective this paper is the determination of $F_{L}$ with
respect to the $F_{2}^{c}$. In this context it is interesting to
recall that $F_{L}$ and $F_{k}^{c}(k=2,L)$ contain an appreciable
part directly sensitive to the gluon density at small $x$. In
addition, we systematically analyze the relation between this
approach and charm structure functions $F_{k}^{c}(k=2,L)$, as
these results are independent of the
exact gluon kinematics .\\

Let us first discuss  charm production, and some phenomenological
aspects of the observable relevant to the charm structure
functions for the experimental data at HERA,  and its contribution
 to the longitudinal structure function $F_{L}$. In the case of heavy quark production, we can have
condition that the heavy quarks produced from the boson- gluon
fusion (BGF) via $\gamma^{*}g{\rightarrow}c\overline{c}$. That is,
in PQCD calculations the production of heavy quarks at HERA
proceeds dominantly via the direct BGF where the photon interacts
with a gluon in the proton by the exchange of a heavy quark pair
[5-12]. In NLO perturbative QCD, the charm structure functions
$F_{k}^{c}(k=2,L)$ are given by [13]
\begin{eqnarray}
F_{k}^{c}(x,Q^{2})&=&C_{g,k}^{c}(\frac{x}{y},Q^{2}){\otimes}G(y,Q^{2})\\\nonumber
&&=2xe_{c}^{2}\frac{\alpha_{s}(\mu^{2})}{2\pi}\int_{ax}^{1}\frac{dy}{y}C_{g,k}^{c}
(\frac{x}{y},\zeta)g(y,\mu^{2}),
\end{eqnarray}
where $a=1+4\zeta(\zeta{\equiv}\frac{m_{c}^{2}}{Q^{2}})$ and the
renormalization scale $\mu$ is assumed to be either
$\mu^{2}=4m_{c}^{2}$ or $\mu^{2}=4m_{c}^{2}+Q^{2}$. Here
$C^{c}_{g,k}$ is the charm coefficient function in LO and NLO
analysis as
\begin{eqnarray}
C_{k,g}(z,\zeta)&{\rightarrow}&C^{0}_{k,g}(z,\zeta)+a_{s}(\mu^{2})[C_{k,g}^{1}(z,\zeta)\\\nonumber
&&+\overline{C}_{k,g}^{1}(z,\zeta)ln\frac{\mu^{2}}{m_{c}^{2}}],
\end{eqnarray}
where $a_{s}(\mu^{2})=\frac{\alpha_{s}(\mu^{2})}{4\pi}$ and in the
NLO analysis
\begin{eqnarray}
\alpha_{s}(\mu^{2})=\frac{4{\pi}}{\beta_{0}ln(\mu^{2}/\Lambda^{2})}
-\frac{4\pi\beta_{1}}{\beta_{0}^{3}}\frac{lnln(\mu^{2}/\Lambda^{2})}{ln(\mu^{2}/\Lambda^{2})}
\end{eqnarray}
with $\beta_{0}=11-\frac{2}{3}n_{f},
\beta_{1}=102-\frac{38}{3}n_{f} $ ($n_{f}$ is the number of active
flavours), $\zeta{\equiv}\frac{m_{c}^{2}}{Q^{2}}$ and  $\mu$ is
the renormalization scale.\\
In the LO analysis, the coefficient functions BGF can be found
[14-16], as
\begin{eqnarray}
C^{0}_{g,2}(z,\zeta)&=&\frac{1}{2}([z^{2}+(1-z)^{2}+4z\zeta(1-3z)-8{\zeta^{2}}z^{2}]\nonumber\\
&&{\times}ln\frac{1+\beta}{1-\beta}+{\beta}[-1+8z(1-z)\nonumber\\
&&-4z{\zeta}(1-z)]),
\end{eqnarray}
and
\begin{eqnarray}
C^{0}_{g,L}(z,\zeta)=-4z^{2}{\zeta}ln\frac{1+\beta}{1-\beta}+2{\beta}z(1-z),
\end{eqnarray}
where $\beta^{2}=1-\frac{4z\zeta}{1-z}$. At NLO,
$O(\alpha_{em}\alpha_{s}^{2})$, the contribution of the photon-
gluon component is usually presented in terms of the coefficient
functions $C_{k,g}^{1}, \overline{C}_{k,g}^{1}$. Using the fact
that  the virtual photon- quark(antiquark) fusion subprocess can
be neglected, because their contributions to the heavy-quark
leptoproduction vanish at LO and are small at NLO.  In a wide
kinematic range, the contributions to the charm structure
functions in NLO are not positive due to mass factorization.
Therefore the charm structure functions are dependence to the
gluonic observable in LO and NLO. The NLO coefficient functions
are only available as computer codes[13,17]. But in the high-
energy regime ($\zeta<<1$) we can used the compact form of these
coefficients according to
the Refs.[18,19].\\

Using Eq.6, We recall the relation between the charm structure
functions and the gluon distribution function at small $x$  and
the limit $\mu^{2}{\rightarrow} Q^{2}$, we obtain the effective
formula for the charm structure functions as
\begin{eqnarray}
F_{k}^{c}(x,Q^{2})&=&G(x,Q^{2})[C_{g,k}^{c}(1-z,Q^{2}){\otimes}(1-z)^{\delta}]\nonumber\\
&&=2e_{c}^{2}\frac{\alpha_{s}(\mu^{2})}{2\pi}G(x,Q^{2}){\times}\nonumber\\
&&\int_{1-\frac{1}{a}}^{1-x}C_{g,k}^{c}
(1-z,\zeta)(1-z)^{\lambda_{g}}dz,
\end{eqnarray}
 here $C_{g,k}^{c}$ is defined by Eq.9. The main input to the Eqs.7 and 13 is  the gluon distribution
$G(x,Q^{2})$. These equations give predictions  for the structure
functions as a function of the gluon distribution. Inserting Eq.7
in Eq.13, we obtain our master formula for the charm structure
functions into the longitudinal proton structure function, as
\begin{equation}
F_{k}^{c}(x,Q^{2})=\frac{[C_{g,k}^{c}(1-z,Q^{2}){\otimes}(1-z)^{\delta}]}{[K^{G}(1-z,Q^{2}){\otimes}(1-z)^{\delta}]}F^{g}_{L}(x,Q^{2}).
\end{equation}
 In fact, this equation which is independent of the gluon distribution function
 , is very useful for practical
 applications. In this equation we used the solutions of the NLO
 BGF analysis ($C_{g,k}^{c}$), NLO kernels ($K^{G}$)
  and considered $\delta$ as a hard (Lipatov) Pomeron exponent. We observe that the longitudinal structure function measurement
   at DESY $ep$ collider HERA will be able to make a reasonably precise measurement of
 the charm structure functions at low $x$ values.\\

 Having checked that this formula reproduces satisfactory the
 existing the longitudinal proton structure function  into the
 charm structure function at high $Q^{2}$ values. One finds the following final form for the longitudinal structure
 function, as
 \begin{equation}
F^{g}_{L}(x,Q^{2})=\frac{[K^{G}(1-z,Q^{2}){\otimes}(1-z)^{\delta}]}
{[C_{g,k}^{c}(1-z,Q^{2}){\otimes}(1-z)^{\delta}]}F_{2}^{c}(x,Q^{2}).
\end{equation}
Recently, H1 Collaboration measures the longitudinal proton
structure function [11] and charm structure function [12], as data
for the longitudinal proton structure function are existing only
at $Q^{2}{\leq}45 GeV^{2}$. Therefore, when analysis the charm
structure function, it is particularly important to obtain our
prediction according to Eq.15 for the longitudinal  structure
function at $Q^{2}{>}45 GeV^{2}$, rather than to low $Q^{2}$
values of
$F_{L}(x,Q^{2})$. \\

We start our numerical analysis by comparing the calculations of
the structure functions with the experimental data and other
models. To be precise, we use the formulae (14,15) with
$\Lambda=0.224 GeV$,
 $m_{c}=1.25GeV$ and $\delta{\simeq}0.5$. Now
extract $F_{k}^{c\overline{c}} $ from the H1 measurements of the
longitudinal proton structure function [11] in Eq.14. Our NLO
results for the charm structure functions
$F_{k}^{^{c}}(x,Q^{2})(k=2\hspace{0.1cm} \& \hspace{0.1cm}L)$ are
presented in Table 1 for each bin in $Q^{2}$ and $x$. In Table 2
we show our results for the charm structure functions at $Q^{2}=20
GeV^{2} $ for various values of $x$. In Figs.1 and 2 we show the
calculations obtained  for the charm structure functions using
this approach with the longitudinal structure function data. The
charm structure functions are plotted as a function of $x$ at
$Q^{2}=20 GeV^{2}$. The data shown in Fig.1 are from H1 [12]
experiment. In Figs.1 and 2 we compared our results for the
$F_{2}^{c}$ and $F_{L}^{c}$
 with  DL fit [7], GJR parameterization [20] and
color-dipole model (CDM) [21], respectively. The agreement between
the experimental data and other models with our calculations is
good.\\

We stress though the importance of the existent data in the
evaluation of the longitudinal structure function at high $Q^{2}$
values. In Table 3 we predict the longitudinal structure function
$F_{L}(x,Q^{2})$ with respect to the H1 measurements of the charm
structure function $F_{2}^{^{c}}(x,Q^{2})$ [12] in Eq.15 at low
and high $Q^{2}$ values. Our NLO results for $Q^{2}=20, 60, 200$
and $650 GeV^{2}$ are presented. We observe that theoretical
uncertainty related to the freedom in the choice of the
renormalization scales $\mu^{2}=4m^{2}_{c}$ and
$\mu^{2}=4m^{2}_{c}+Q^{2}$ as accompanied to the total errors of
the measurements data at the quadrature procedure. In Fig.3, we
show the prediction of Eq.15 for the longitudinal structure
function. We compare our results for $F_{L}$ with the H1 [11] data
and DL model [22]. As can be seen, the values of the longitudinal
structure function increase as $x$ decreases. This is because the
hard- Pomeron exchange defined by
DL model is expected to hold in the small- $x$ limit.\\

In summary, we have presented Eq.15 for the extraction of the
longitudinal structure function $F_{L}$ at low $x$ and high
$Q^{2}$ values from the charm structure function $F_{2}^{c}$. This
approximation relation provide the possibility the non-direct
determination $F_{L}$. This is important since the direct
extraction of $F_{L}$ from experimental data is a cumbersome
procedure. We have found that the charm structure function gave us
the longitudinal structure function that agrees well with the
phenomenological fit.  Having checked that this approach gives a
good description of the data, we have used it to predict  $F_{L}$
and $F_{2}^{c}$ to be measured in collisions.\\

\newpage
\subsection{Appendix}
The explicit form of the gluon kernel is given by the following
from:
\begin{widetext}
\begin{eqnarray}
K^{G}(\frac{x}{y},Q^{2})&=&\frac{\alpha_{s}}{4\pi}[8(x/y)^{2}(1-x/y)][\sum_{i=1}^{N_{f}}e_{i}^{2}]
+(\frac{\alpha_{s}}{4\pi})^{2}[\sum_{i=1}^{N_{f}}e_{i}^{2}]16C_{A}(x/y)^2(+4dilog(1-x/y)\nonumber\\
&&-2(1-x/y)ln(x/y)ln(1-x/y)+2(1+x/y)dilog(1+x/y)+3ln(x/y)^2
+2(x/y-2)\pi^2/6\nonumber\\
&&+(1-x/y)ln(1-x/y)^2+2(1+x/y)ln(x/y)ln(1+x/y)
+\frac{(24+192x/y-317(x/y)^2)}{24(x/y)}ln(x/y)\nonumber\\
&&+\frac{(1-3x/y-27(x/y)^2+29(x/y)^3)}{3(x/y)^2}ln(1-x/y)
+\frac{(-8+24x/y+510(x/y)^2-517(x/y)^3)}{72(x/y)^2}\nonumber\\
&&-16C_{F}(x/y)^2(\frac{5+12(x/y)^2}{30}ln(x/y)^2
-(1-x/y)ln(1-x/y)+\frac{-2+10(x/y)^3-12(x/y)^5)}{15(x/y)^3}\nonumber\\
&&(+dilog(1+x/y)+ln(x/y)ln(1+x/y))+2\frac{5-6(x/y)^2}{15}\pi^2/6+\frac{4-2x/y-27(x/y)^2-6(x/y)^3}{30(x/y)^2}ln(x/y)\nonumber\\
&&+\frac{(1-x/y)(-4-18x/y+105(x/y)^2)}{30(x/y)^2}).\nonumber\\
\end{eqnarray}
\end{widetext}
For the SU(N) gauge group, we have $C_{A}=N$,
$C_{F}=(N^{2}-1)/2N$,
 $T_{F}=n_{f}T_{R}$, and $T_{R}=1/2$ where $C_{F}$ and $C_{A}$ are the color Cassimir operators.

 %%%%%%%%%%%%%%%%%%%%%%%%%%%%%%%%%%%%%%%%%%%%%%%%%%%%%%%
\subsection{Acknowledgments}
G.R.Boroun thanks Prof.A.Cooper-Sarkar for interesting and useful
discussions.\\

%%%%%%%%%%%%%%%%%%%%%%%%%%%%%%%%%%%%%%%%%%%%%%%%%%%%%%%%%%%%%%%%%%%%%%%%
\newpage
\textbf{References}\\
1.A.M.Cooper-Sarkar, et.al., Z.Phys.C\textbf{39},
 281(1988); A.M.Cooper-Sarkar and R.C.E.Devenish, Acta.Phys.Polon.B\textbf{34},
 2911(2003).\\
2.D.I.Kazakov, et.al., Phys.Rev.Lett\textbf{65}, 1535(1990).\\
3.J.L.Miramontes, J.sanchez Guillen and E.Zas, Phys.Rev.D \textbf{35}, 863(1987).\\
4.C.Lopez and F.J.Yndurain, Nucl.Phys.B \textbf{171}, 231(1980);
\textbf{183}, 157(1981); A.V.Kotikov,Phys.Rev.D \textbf{49},
5746(1994); A.V.Kotikov, Phys.Atom.Nucl.\textbf{
59}, 2137(1996).\\
5. A.Vogt, arXiv:hep-ph:9601352v2(1996).\\
6. H.L.Lai and W.K.Tung, Z.Phys.C\textbf{74},463(1997).\\
7. A.Donnachie and P.V.Landshoff, Phys.Lett.B\textbf{470},243(1999).\\
8. N.Ya.Ivanov, Nucl.Phys.B\textbf{814}, 142(2009); N.Ya.Ivanov
and B.A.Kniehl, Eur.Phys.J.C\textbf{59}, 647(2009).\\
9. F.Carvalho, et.al., Phys.Rev.C\textbf{79}, 035211(2009).\\
10. S.J.Brodsky, P.Hoyer, C.Peterson and
N.Sakai,Phys.Lett.B\textbf{93}, 451(1980); S.J.Brodsky, C.Peterson
and N.Sakai, Phys.Rev.D\textbf{23}, 2745(1981).\\
11.F.D. Aaron et al. [H1
Collaboration],Eur.Phys.J.C\textbf{71},1579(2011).\\
12. F.D. Aaron et al. [H1
Collaboration],Eur.Phys.J.C\textbf{65},89(2010).\\
13.M.Gluk, E.Reya and A.Vogt, Z.Phys.C\textbf{67}, 433(1995); Eur.Phys.J.C\textbf{5}, 461(1998).\\
14.V.N. Baier et al., Sov. Phys. JETP 23 104 (1966); V.G. Zima,
Yad. Fiz. 16 1051 (1972); V.M. Budnev et al., Phys. Rept. 15 181
(1974).\\
15.E. Witten, Nucl. Phys. B104 445 (1976); J.P. Leveille and T.J.
Weiler, Nucl. Phys. B147 147 (1979); V.A. Novikov et al., Nucl.
Phys. B136 125 (1978) 125.\\
16.E. Witten, Nucl. Phys. B104 445 (1976); J.P. Leveille and T.J.
Weiler, Nucl. Phys. B147 147 (1979); V.A. Novikov et al., Nucl.
Phys. B136 125 (1978) 125.\\
17.E.Laenen, S.Riemersma, J.Smith and W.L. van Neerven,
Nucl.Phys.B \textbf{392}, 162(1993).\\
18. A.~Y.~Illarionov,B.~A.~Kniehl and A.~V.~Kotikov, Phys.Lett. B {\bf 663}, 66 (2008).\\
19. S. Catani, M. Ciafaloni and F. Hautmann, Preprint
CERN-Th.6398/92, in Proceeding of the Workshop on Physics at HERA
(Hamburg, 1991), Vol. 2., p. 690; S. Catani and F. Hautmann, Nucl.
Phys. B \textbf{427}, 475(1994); S. Riemersma, J. Smith and W. L.
van Neerven, Phys. Lett. B \textbf{347}, 143(1995).\\
20.M. Gluck, P. Jimenez-Delgado, E. Reya,
Eur.Phys.J.C\textbf{53},355(2008).\\
21. N.N.Nikolaev and V.R.Zoller, Phys.Lett. B\textbf{509},
283(2001).\\
22. A.Donnachie and P.V.Landshoff, Phys.Lett.B\textbf{550}, 160(2002).\\

%%%%%%%%%%%%%%%%%%%%%%%%%%%%%%%%%%%%%%%%%%%%%%%%%%%%%%%%%%%%%%%
\begin{table}[h]
\centering \caption{Predictions of the charm structure functions
$F_{k}^{c}$ from the averaging longitudinal proton structure
function that accompanied with the total errors (Table 23 at
Ref.[11]). The uncertainties in our results associated with the
renormalization scales $\mu^{2}=4m_{c}^{2}$ and
$\mu^{2}=4m_{c}^{2}+Q^{2}$ and the longitudinal structure function
total error.}\label{table:table2}
\begin{minipage}{\linewidth}
\renewcommand{\thefootnote}{\thempfootnote}
\centering
\begin{tabular}{|l|c||c|c||c|c||l|l|} \hline\noalign{\smallskip} $Q^{2}(GeV^{2})$ & $ <x>$ &$
<F_{L}>$ & ${\Delta}F_{L}$ & $F_{2}^{c}$ & ${\Delta}F_{2}^{c}$& $F_{L}^{c}$ & ${\Delta}F_{L}^{c}$ \\
\hline\noalign{\smallskip}
12 & 0.000319 & 0.314 & 0.058 & 0.203& 0.072 & 0.038 & 0.059\\
15 & 0.000402 & 0.255 & 0.058 & 0.198 & 0.076 & 0.040 & 0.059\\
20 & 0.000540 & 0.312 & 0.061 & 0.310 & 0.110 & 0.070 & 0.066\\
25 & 0.000686 & 0.269 & 0.069 & 0.313 & 0.126 & 0.074 & 0.076\\
35 & 0.00103 & 0.201 & 0.082 & 0.294  & 0.143  & 0.075 & 0.090\\
45 & 0.00146 & 0.219 & 0.116 & 0.376  & 0.202  & 0.101 & 0.128\\
\hline\noalign{\smallskip}
\end{tabular}
\end{minipage}
\end{table}
\begin{table}[h]
\centering \caption{Predictions of the charm structure functions
$F_{k}^{c}$ from the longitudinal proton structure function that
accompanied with the total errors (Table 22 at Ref.[11]) at
$Q^{2}=20 GeV^{2}$ . The uncertainties in our results associated
with the renormalization scales $\mu^{2}=4m_{c}^{2}$ and
$\mu^{2}=4m_{c}^{2}+Q^{2}$ and the longitudinal structure function
total error. }\label{table:table3}
\begin{minipage}{\linewidth}
\renewcommand{\thefootnote}{\thempfootnote}
\centering
\begin{tabular}{|l||c|c||c|c||c|l|} \hline\noalign{\smallskip} $x$ & $F_{L}$ &$
{\delta}F_{L}$ & $F_{2}^{c}$ & ${\delta}F_{2}^{c}$& $F_{L}^{c}$& ${\delta}F_{L}^{c}$ \\
\hline\noalign{\smallskip}
0.372E-3 & 0.209 & 0.238 & 0.206 & 0.246& 0.047 & 0.239 \\
0.415E-3 & 0.309 & 0.189 & 0.305 & 0.193& 0.068 & 0.189 \\
0.464E-3 & 0.402 & 0.135 & 0.396 & 0.142& 0.088 & 0.136 \\
0.526E-3 & 0.347 & 0.122 & 0.241 & 0.127& 0.076 & 0.122 \\
0.607E-3 & 0.289 & 0.141 & 0.285 & 0.145& 0.064 & 0.141 \\
0.805E-3 & 0.194 & 0.188 & 0.191 & 0.189& 0.043 & 0.188 \\
\hline\noalign{\smallskip}
\end{tabular}
\end{minipage}
\end{table}
\begin{table}[h]
\centering \caption{Predictions of the longitudinal proton
structure function from the charm structure function $F_{2}^{c}$
that accompanied with the total errors [12] at $Q^{2}=20, 60, 200$
and $650 GeV^{2}$ . The uncertainties in our results associated
with the renormalization scales $\mu^{2}=4m_{c}^{2}$ and
$\mu^{2}=4m_{c}^{2}+Q^{2}$ and the charm structure function total
error. }\label{table:table3}
\begin{minipage}{\linewidth}
\renewcommand{\thefootnote}{\thempfootnote}
\centering
\begin{tabular}{|l|c|c||c|c||c|l|} \hline\noalign{\smallskip} $Q^{2}$ & $x$ &$
y$ & $F_{2}^{c}$ & ${\delta}F_{2}^{c}$& $F_{L}$& ${\delta}F_{L}$ \\
\hline\noalign{\smallskip}
20& 0.002 & 0.098 & 0.188 & 0.011& 0.211 & 0.064 \\
20& 0.0013 & 0.151 & 0.219 & 0.011& 0.245 & 0.074 \\
20& 0.0008 & 0.246 & 0.276 & 0.011& 0.308 & 0.093 \\
20& 0.0005 & 0.394 & 0.287 & 0.010& 0.320 & 0.095 \\
60& 0.005 & 0.118 & 0.199 & 0.011& 0.130 & 0.065 \\
60& 0.0032 & 0.185 & 0.264 & 0.011& 0.171 & 0.085 \\
60& 0.002 & 0.295 & 0.339 & 0.010& 0.220 & 0.105 \\
60& 0.0013 & 0.454 & 0.307 & 0.010& 0.198 & 0.098 \\
200& 0.013 & 0.151 & 0.160 & 0.027& 0.079 & 0.059 \\
200& 0.005 & 0.394 & 0.243 & 0.029& 0.118 & 0.083 \\
650& 0.032 & 0.200 & 0.085 & 0.034& 0.038 & 0.045 \\
650& 0.013 & 0.492 & 0.203 & 0.033& 0.088 & 0.075 \\
\hline\noalign{\smallskip}
\end{tabular}
\end{minipage}
\end{table}
%%%%%%%%%%%%%%%%%%%%%%%%%%%%%%%%%%%%%%%%%%%%%%%%%%%%%%
\begin{figure}
\includegraphics{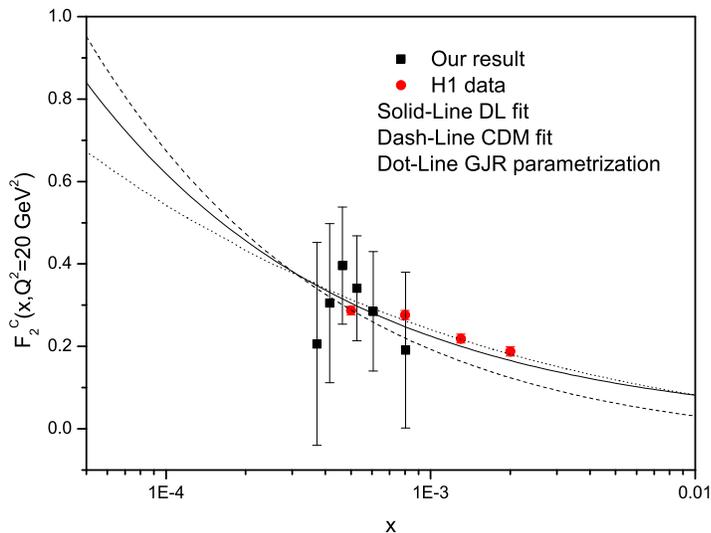}
\caption{The charm component of the structure function at
$Q^{2}=20 GeV^{2}$ according to the longitudinal structure
function input [11]. Our results accompanied with the errors due
to the renormalization scales, compared to H1 data [12], and also
A.Donnachie- P.V.Landshoff (DL) model [7], GJR parameterization
[20] and color dipole model (CDM) [21] . }\label{Fig1}
\end{figure}
\begin{figure}
\includegraphics{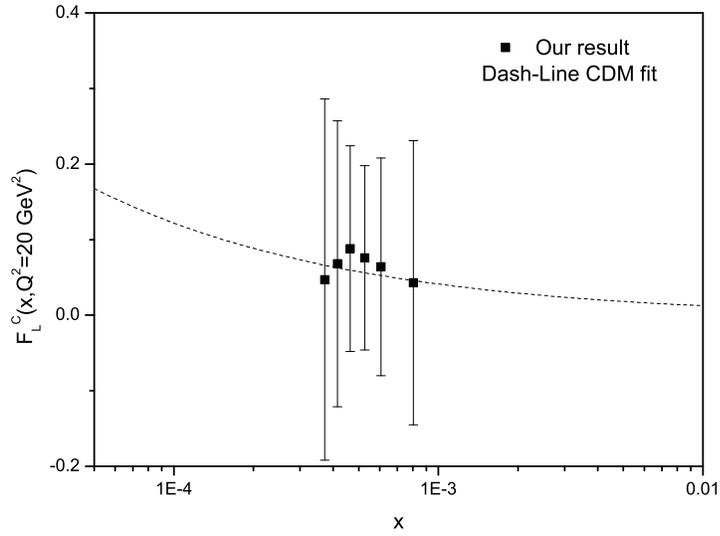}
\caption{The charm component of the longitudinal structure
function at $Q^{2}=20 GeV^{2}$ according to the longitudinal
structure function input [11]. Our results accompanied with the
errors due to the renormalization scales, compared only to the
color dipole model (CDM) [21]. }\label{Fig2}
\end{figure}
\begin{figure}
\includegraphics[width=1\textwidth]{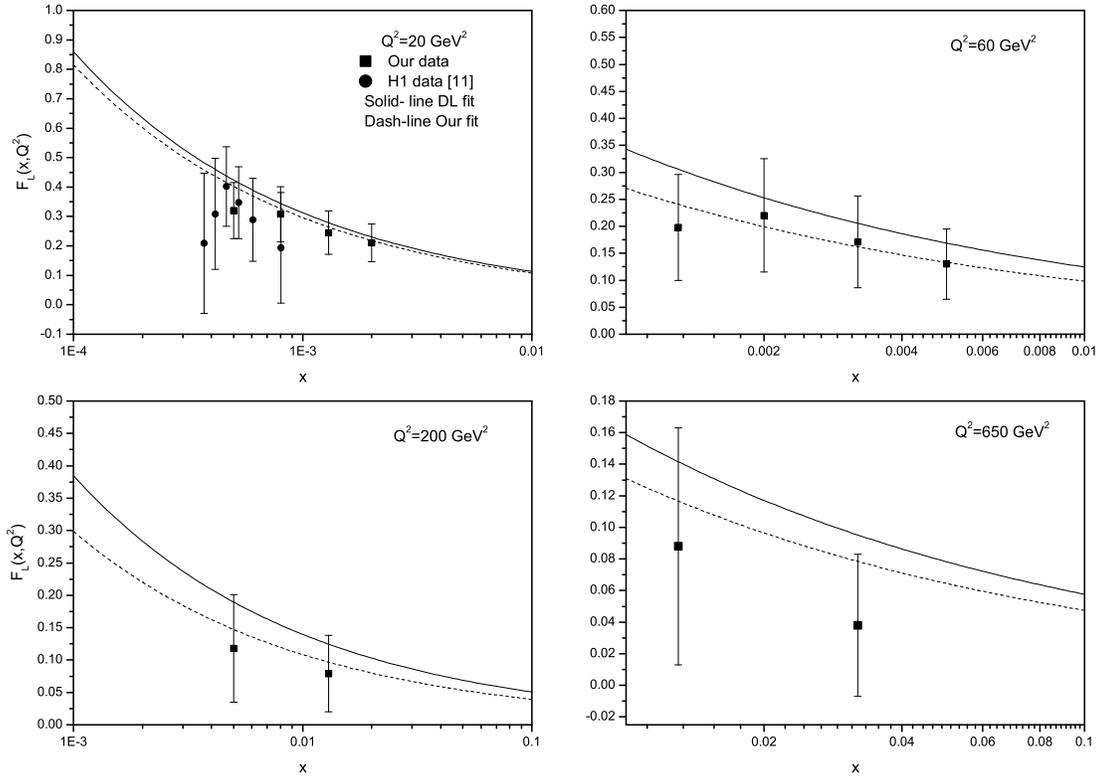}
\caption{Predictions for $F_{L}(x,Q^{2})$ at NLO, from the charm
structure function data [12] at $Q^{2}$ values of 20, 60, 200 and
650 $GeV^{2}$. Our results accompanied with the errors due to the
renormalization scales, compared to the DL model [22].
}\label{Fig3}
\end{figure}
%%%%%%%%%%%%%%%%%%%%%%%%%%%%%%%%%%%%%%%%%%%%%%%%%%%%%%%%%%%%
\end{document}